\documentclass{emulateapj}

\shorttitle{[OIII] blue outliers}
\shortauthors{Komossa et al.}

\begin{document}

\def\approxgt{\mathrel{\hbox{\rlap{\lower.55ex \hbox {$\sim$}}
        \kern-.3em \raise.4ex \hbox{$>$}}}}

\title{On the nature of Seyfert galaxies with high [OIII]5007 blueshifts} 
\author{S. Komossa}
\affil{Max-Planck-Institut f\"ur extraterrestrische Physik,
Giessenbachstrasse 1, 85748 Garching, Germany; skomossa@mpe.mpg.de}

\author{D. Xu}
\affil{National Astronomical Observatories, Chinese Academy of Science,
A20 Datun Road, Chaoyang District,
Beijing 100012, China}

\author{H. Zhou}
\affil{Max-Planck-Institut f\"ur extraterrestrische Physik,
Giessenbachstrasse 1, 85748 Garching, Germany}

\author{T. Storchi-Bergmann}
\affil{Instituto de Fisica, UFRGS, Campus do Vale, CP 15051, Porto Alegre 91501-970, RS, Brazil}

\and 

\author{L. Binette}
\affil{D{\'e}partement de Physique, de G{\'e}nie Physique et d'Optique, Universit{\'e}
Laval, Qu{\'e}bec, QC, G1K7P4, Canada; and Instituto de Astronomia, 
UNAM, Ap. 70-264, 04510 M{\'e}xico, DF, M{\'e}xico }

\begin{abstract}
We have studied the properties of Seyfert galaxies with high [OIII]5007 blueshifts
(``blue outliers''), originally identified because of their strong
deviation from the $M_{\rm BH}-\sigma$ relation of normal, narrow-line Seyfert 1 (NLS1) and 
broad-line Seyfert 1 (BLS1) galaxies.
These blue outliers turn out to be important test-beds for 
models of the narrow-line region (NLR), for mechanisms of driving large-scale outflows,
for links between NLS1 galaxies and radio galaxies, and for orientation-dependent
NLS1 models.   
We report the detection of a strong correlation of line blueshift with
ionization potential in each galaxy, including the measurement of 
coronal lines with radial velocities up to 500--1000 km\,s$^{-1}$, and we confirm a 
strong correlation between [OIII] blueshift and line width. 
All [OIII] blue outliers have narrow widths of their broad Balmer lines
and high Eddington ratios.  
While the presence of non-shifted low-ionization lines
signifies the presence of a classical outer quiescent NLR
in blue outliers, we also report the 
absence of any second, non-blueshifted [OIII] component from
a classical inner NLR.
These results 
place tight constraints on NLR models. We favor
a scenario in which the NLR clouds are entrained in a decelerating wind
which explains the strong stratification and the absence of a zero-blueshift
inner NLR of blue outliers. 
The origin of the wind remains speculative at this time 
(collimated radio plasma, thermal winds, radiatively accelerated clouds).
It is perhaps linked to
the high Eddington ratios of blue outliers.  Similar, less powerful winds
could be present in all Seyfert galaxies, but would generally only affect the 
coronal line region (CLR), 
or level off even before reaching the CLR. 
Similarities between blue outliers in NLS1 galaxies and (compact) radio sources 
are briefly discussed. 
\end{abstract}

\keywords{galaxies: active -- galaxies: evolution -- galaxies: individual (SBS0919+515,
SDSSJ115533.50+010730.4, RXJ01354-0043, NGC450\#86, SDSSJ032606.75+011429.9, IRAS11598-0112, 
SDSSJ171828.99+573422.3, PG1244+026, RXJ09132+3658) -- 
galaxies: Seyfert -- quasars: emission lines}

\section{Introduction}

The concept of feedback due to outflows  
is a potential 
key ingredient in understanding   
the coevolution of galaxies and black holes 
(Silk \& Rees 1998, Fabian 1999, Wyithe \& Loeb 2003).
Recent analytic estimates and
simulations demonstrate the importance of feedback from winds/outflows
for instance in cosmic downsizing
(Scannapieco et al. 2005),  
in fixing the $M_{\rm BH}-\sigma$ relation (di Matteo et al. 2005),
and in determining galaxy colors (Springel et al. 2005)
by regulating star formation
in the host galaxy. 
Powerful gaseous outflows in Active Galactic Nuclei (AGN)  
deposit mass, energy and metals
in the interstellar medium of the galaxy and the 
intergalactic medium or intracluster medium (Colbert et al. 1996, Churazov et al. 2001,
Moll et al. 2007). 
AGN winds also play a potentially important role in
unified models of AGN (Elvis 2000, 2006). 

Observational evidence for outflows in AGN exists on various
scales and in all wavebands from the radio
(Morganti et al. 2005, Gallimore et al. 2006), to the IR (Rodriguez-Ardila et al. 2006),
optical (Das et al. 2005), UV (Sulentic et al. 2007, Rodriguez Hidalgo et al. 2007,
Crenshaw \& Kraemer 2007),
and X-ray band (Chelouche \& Netzer 2005, Krongold et al. 2007); 
see Veilleux et al. (2005) for a review. 
There are several lines of evidence that outflows are
particularly strong in narrow-line Seyfert 1 (NLS1) galaxies.
NLS1 galaxies are a subclass of AGN with extreme
emission-line and continuum properties 
which
appear to be in part driven by
their high Eddington ratios, low black hole
masses, 
and
other parameters (e.g., Osterbrock \& Pogge 1985, Boroson 2002,
Grupe 2004; see Komossa 2008
for a review).  The high Eddington ratios likely lead to strong,
radiation-pressure driven outflows. 

As a class of objects with low black hole masses,
high accretion rates and strong winds, the location of NLS1 galaxies 
on the $M_{\rm BH}-\sigma$ plane is of special
interest (Mathur et al. 2001; see our Sect. 4.1).
Komossa \& Xu (2007) have shown that 
NLS1 
galaxies follow the same $M_{\rm BH}-\sigma$ relation
as normal and broad-line Seyfert 1 (BLS1) galaxies, 
if the width of the [SII]6716,6731
emission lines is used 
as a surrogate for stellar velocity dispersion. 
The width of [OIII]5007 (after removal of asymmetric blue
wings) is still a good proxy for stellar velocity
dispersion in BLS1 and NLS1 galaxies, with one important
exception. A subset of NLS1 galaxies    
deviates systematically from the $M_{\rm BH}-\sigma$ relation, 
and all of these are  
characterized by high [OIII] {\em blueshifts}.
At the same time, the [SII]-based measurements
of velocity dispersion of the same objects still place
them {\em on} the $M_{\rm BH}-\sigma$ relation (Fig. 1).
Therefore, the [OIII] lines of these particular galaxies 
were not suitable to estimate $\sigma$. 
However, the independent question is raised as to
which mechanism drives these [OIII] line blueshifts, and what we can learn
from them regarding the nature of these systems 
and perhaps
the evolutionary state of NLS1 galaxies.  That
is the topic of this study.  

A measurable difference in blueshifts and widths of 
different NLR emission lines has been recognized
early (review by Osterbrock 1991). 
The phenomenon of strong [OIII] blueshifts
exceeding one--several hundred km\,s$^{-1}$ (so-called blue outliers),
is more rare. It has been detected early in individual objects (Phillips 1976) 
but has only been studied systematically
recently
(Zamanov et al. 2002, Marziani et al. 2003, Aoki et al. 2005,
Boroson 2005, Bian et al. 2005).
These studies have shown that blue outliers
have high Eddington ratios and small Balmer line widths
(all of them with FWHM(H$\beta$) $<$ 4000 km\,s$^{-1}$, and most of them with
FWHM(H$\beta$) $<$ 2000 km\,s$^{-1}$; Marziani et al. 2003).
It has remained uncertain whether there is (Bian et al. 2005)
or is not (Aoki et al. 2005) a direct correlation between 
[OIII] blueshift and Eddington ratio.
Common to most studies is the presence of a strong correlation between
[OIII] blueshift and [OIII] line width. 
This correlation is also known
among iron coronal lines of BLS1 galaxies (Penston et al. 1984, Erkens et al. 1997).
The reason is not yet well understood. Outflows are
thought to play a role in explaining the phenomenon.

Except Boroson (2005) all sample studies concentrated
on the [OIII]-H$\beta$ region of optical
spectra of blue outliers in order to examine
the phenomenon, while for the first time we include
information from all the detected optical NLR lines. 
We report the detection of strong correlations
and discuss consequences for the nature of blue outliers
and for dynamical NLR models. 
This paper is organized as follows. In Sect. 2 we
describe the sample selection and provide
details on the data analysis.  
Results on
trends and correlations are presented in Sect. 3 which are
then discussed in Sect. 4. A summary and conclusions are
given in Sect. 5. Some individual objects turn out to be remarkable.
Notes on them are 
provided in an Appendix. 
We use a cosmology with 
$H_{\rm 0}$=70 km\,s$^{-1}$\,Mpc$^{-1}$, $\Omega_{\rm M}$=0.3
and $\Omega_{\rm \Lambda}$=0.7 throughout this paper.

\section{Data analysis}

\subsection{The sample} 

The nine NLS1 galaxies of the present work
were drawn from the sample 
of Xu et al. (2007, 2008). 
The original sample selection
and standard data reduction procedures were
described in detail in that work. In brief, 
the sample consists of
NLS1 galaxies from the catalogue of V{\'e}ron-Cetty \& V{\'e}ron
(2003) and a comparison sample of BLS1 galaxies
from Boroson (2003) at redshift $z < 0.3$, which have
Sloan Digital Sky Survey (SDSS)
DR3 (Abazajian et al. 2005) spectra
available and which have 
detectable low-ionization lines
(presence of [SII]6716,6731, with S/N$>$5).  
The BLS1 and NLS1 galaxy samples have similar
redshift and absolute
magnitude distributions.
Xu et al. (2007) corrected the  SDSS spectra
for Galactic extinction, decomposed
the continuum into host galaxy and AGN components, and then subtracted
the host galaxy contribution 
and the FeII complexes from the spectra. 
Emission line profiles of the galaxies were fit with Gaussians
using the IRAF package SPECFIT (Kriss 1994).
Measured FWHMs were corrected for instrumental
broadening.
Re-classification after spectral emission-line fitting led to
39 BLS1 and 55 NLS1 galaxies in the sample. 
We focus here on the nine galaxies with the highest
blueshifts of [OIII] (Tab. 1) which were
identified by Komossa \& Xu (2007; KX07 hereafter). 
Results from the complete NLS1 (and BLS1) sample 
are shown for comparison
purposes, and in order to identify trends
across the whole BLS1 -- NLS1 -- blue-outlier population.  
When we report measurements of optical FeII strength (FeII4570), this is
the integrated flux of the FeII emission complex between
the rest wavelengths 4434\AA\ and 4684\AA.
We refer to the flux sum of the two Sulphur lines, [SII]6716 and [SII]6731,
as [SII]6725.  

\subsection{Emission-line fits and [OIII] profile}

The Balmer lines were 
decomposed into three components, narrow core (H$\beta_{\rm n}$; FWHM fixed to
that determined for [SII]6716,6731), and two broad components.
No physical meaning is ascribed to the two separate
broad components; they merely serve as a mathematical
description (see Xu et al. 2008 for alternative
Lorentzian fits). The final width of the broad-line emission, 
H$\beta_{\rm b}$, is determined
as the FWHM of the sum of the two Gaussians.  
With the exception
of [OIII], forbidden lines are well represented by
single Gauss profiles. The total [OIII] emission-line profile, [OIII]$_{\rm totl}$,
was decomposed into two Gaussian components,
a narrow core ([OIII]$_{\rm c}$) and broad base. 
We distinguish between two types of [OIII]
spectral complexity: (1) the presence of such a broad base which tends 
to be blue-asymmetric (e.g., Heckman et al. 1981), 
and is referred to as ``blue wing'', and (2) 
systematic blueshifts of the whole core of [OIII]. Objects which show
this latter phenomenon are 
called ``blue outliers'' (Zamanov et al. 2002){\footnote{Note that we use
a velocity shift of $v_{\rm[OIII]_c}>$150 km\,s$^{-1}$ in order to refer to an object
as [OIII] blue outlier; while Zamanov et al. (2002) use this terminology
for objects with $v_{\rm[OIII]}>$250 km\,s$^{-1}$, where [OIII] outflow velocity is
measured relative to H$\beta$.}}.  
Measurements of the FWHM and blueshift of [OIII] reported in this work
refer to the core of the emission-line, unless noted otherwise.  

Ideally, measurements of the velocity shift of [OIII] should be done
relative to the galaxy restframe, as defined by stellar absorption lines
from the host galaxy. However, these features are generally weak or absent in
our galaxies (and in NLS1 galaxies in general).
We therefore measure
velocity shifts of [OIII] and of all other lines relative to [SII]. 
We use positive velocity values to refer to blueshifts,
negative ones for redshift. The shifts of H$\beta$ and other
low-ionization lines ([OI]6300, NII[6584], [OII]3727) 
agree well with [SII], while other high-ionization
lines ([NeIII]3861, [NeV]3426, and iron coronal lines)
are characterized by high blueshifts (Sect. 3.3).  
If lines were too faint to be fit by a Gaussian
of free width, we used fixed width instead (fixed to FWHM([SII])
for low-ionization lines or to FWHM([OIII]$_{\rm c}$) for high-ionization lines)
and determined the central wavelength to measure the blueshift. 
Results are reported in Tab. 1. 
Since the redshifts provided by the SDSS pipeline are determined based on all
strong detected emission lines and are therefore influenced by the
blueshifted [OIII], we have re-measured redshifts based on H$\beta_{\rm n}$. 
It is these redshifts that are given in Tab. 1. 
We have measured the average [OIII]$_{\rm totl}$/[SII]6725
ratios of our BLS1 and NLS1 galaxy sample. We find that, in terms
of this ratio, their NLRs are similar, 
with $<$[OIII]$_{\rm totl}$/[SII]6725$>$=5 in BLS1 galaxies,
and $<$[OIII]$_{\rm totl}$/[SII]6725$>$=4 in NLS1 galaxies.
 
Notes on individual
objects are given in the Appendix.   Here, we briefly comment on the detection
of [FeX]6375 in two blue outliers. In PG1244+026 [FeX] is blended with 
[OI]6365. For decomposition, the width of [OI]6365 was fixed to that of
[OI]6300. The two-component fit then reproduces the expected ratio
of [OI]6300/[OI]6365 $\approx$ 3. [FeX] is blueshifted by 640 km\,s$^{-1}$.  
In RXJ0135$-$0043, [FeX] overlaps with atmospheric O$_2$ which is 
imperfectly corrected for. Therefore, the width of [FeX] was fixed to that
of [OIII]. This gives an outflow velocity of 990 km\,s$^{-1}$. 

\subsection{Robustness of spectral fits: continuum and line decomposition} 

In order to see how much the fitting procedure affects measurements
of [OIII] fluxes, asymmetries and blueshifts, we have repeated
the spectral analysis several times under different conditions and assumptions:
We have fit the host-galaxy-corrected, FeII subtracted spectrum and 
for comparison the original SDSS spectrum without any of these corrections. 
We have re-fit the [OIII] profile in a number of
different ways, and with different Gaussian decompositions.
The [OIII] profile was fit using the red half of the profile only,
or different fractions of the total profile, or only the upper 30\%  
of the profile, or with a single and a double Gaussian profile.    
Fitting the profile with one single Gaussian component maximizes
the blueshift in objects with significant [OIII] blue wings. 
In all other cases, uncertainties in line blueshifts due to non-subtraction
of a host galaxy contribution and due to different line fitting
procedures 
are typically 
much less than 50 km\,s$^{-1}$. 
The blueshift independently
determined from the weaker line [OIII]4959 also agrees with [OIII]5007 
within better than 50 km\,s$^{-1}$.
Four of our galaxies are also in the sample of Boroson (2005). His [OIII]
blueshift measurements agree with ours within typically better than $\pm$10 km\,s$^{-1}$.   
Uncertainties in FWHM of the core of [OIII] are typically 30\%
for [OIII] lines with extra blue wing (smaller in others), and are due
to uncertainties in decomposition. 
We note in passing that the frequency of [OIII] blue wings in blue outliers is 
neither particularly high nor low (see also Boroson 2005).

\section{Results}

\subsection{Constraints on the presence/absence of blueshifted emission lines} 

In order to facilitate a systematic discussion
of NLR models of blue outliers (Sect. 4.3), we 
have investigated whether the non-shifted low-ionization lines have a (faint) 
high-ionization [OIII] counterpart 
and vice versa{\footnote{Note that, apart from [OIII], other high-ionization lines
show blueshifts, too (Sect. 3.3.1), but they are too faint
for multi-component line decompositions. Among the non-shifted
emission lines, [SII] is the strongest unblended line, and we therefore
focus on [SII] for the corresponding estimates in this section.}}. We did this representatively
for the three galaxies with the highest [OIII]  blueshifts (SBS0919+515, SDSSJ11555+0107
and RXJ01354$-$0043),
and for the galaxy with the highest S/N (PG1244+026). 
We have examined the following three questions:

(1) Is there any [OIII] emission at zero blueshift present 
in the spectrum (which could originate from the inner NLR; or be counterpart
to the low-ionization emission lines from the outer NLR) ?  
In order to test this, we have enforced an extra Gaussian line
contribution to [OIII] at zero blueshift
(or at $v_{\rm [OIII]} = 50$ km\,s$^{-1}$, the average
value of our NLS1 sample excluding blue outliers) 
and of fixed  
FWHM([OIII]$_{\rm extra}$)=FWHM([SII]).
We find that, if present, it must be very weak.  
The average ratio [OIII]$_{\rm totl}$/[SII]6725 in our BLS1 sample is
5, while in our NLS1 sample it is 4. This value
of 4 is much higher than the
upper limit on the extra component which was fit to the observed spectra: 
Typically, [OIII]$_{\rm extra}$/[SII]6725 $\approx 0.1-0.4$,
and always, [OIII]$_{\rm extra}$/[SII]6725 $<1$.   
Since the ratio of [OIII]/[SII] of nearby Seyfert galaxies 
generally decreases in dependence of radius
(in models and observations; Komossa \& Schulz 1997, Bennert et al. 2006)
and since there must be an [OIII] contribution from the outer [SII] emitting clouds,
the limits derived on any non-blueshifted [OIII] contribution from the {\em inner} NLR
are very tight. 

(2) Does the blueshifted [OIII] have a blueshifted H$\beta$
counterpart ? I.e., how much highly blueshifted H$\beta$ could be `hidden' in
the H$\beta$ profile ? 
In order to check this, we have re-fit H$\beta$ adding an additional Gaussian
to describe its profile with parameters of this extra component fixed
at $v_{\rm H\beta_{extra}}=v_{\rm [OIII]_c}$, FWHM(H$\beta_{\rm extra}$)=FWHM([OIII]$_{\rm c}$)
and an intensity ratio [OIII]$_{\rm c}$/H$\beta_{\rm extra}$=10.   
We find that such a weak component could generally be hidden in
the H$\beta$ profile.   

(3) Does the blueshifted [OIII] have a blueshifted low-ionization
counterpart in [SII] ? 
This is not the case. For typical {\em spatially averaged} ratios
of [OIII]/[SII]6725 $\approx$4 of our NLS1 sample, blueshifted [SII] lines
would have generally been detectable, but are absent. 
Very faint blueshifted [SII], as it may arise in the {\em inner} NLR, cannot be
excluded with the present data. 

In summary, we find (1) little evidence for zero-blueshift [OIII]
emission (the amount of emission consistent with the spectra likely originates
from the outer parts of the NLR which also emits in [SII] and other
low-ionization lines); (2) an extra blueshifted component
in H$\beta$ -- the expected counterpart to blueshifted [OIII]-- can be 
hidden in the H$\beta$ profile; and (3) there is
no detectable highly blueshifted component in the low-ionization lines. 

\subsection{Estimates of black holes masses, Eddington ratios and NLR densities}

Black hole masses were estimated by using
the radius($R_{\rm BLR}$)-luminosity($L$) relation of BLS1 galaxies
as reported by Kaspi et al. (2005) and the width of H$\beta$.
The luminosities at 5100\AA~were taken from Xu et al. (2008) 
and are based on SDSS g* and r* magnitudes corrected for Galactic extinction.
Since we lack multi-wavelength spectral energy distributions, 
the bolometric luminosity $L_{\rm bol}$ was estimated 
using a standard bolometric correction of
$L_{\rm bol}=9\,\lambda{L_{\rm5100}}$ (Kaspi et al. 2000).
$L_{\rm Edd}$ was calculated from the black hole mass,
according to $L_{\rm Edd} = 1.3\,10^{38}$ M$_{\rm BH}$/M$_{\odot}$ erg/s.
Typical errors in individual BH mass estimates can be as large as 0.5 dex,
and arise from the use of single-epoch data,
and uncertainties in H$\beta$ decomposition 
and in host galaxy contribution.
The NLR density was derived from the density-sensitive [SII]
intensity ratio,
[SII]6716/[SII]6731
(Xu et al. 2007).

\subsection{Trends and correlations}

We have checked for (1) trends of outflow velocity with
other (atomic) parameters within each single spectrum,
and (2) for trends across our sample of blue outliers
(do all blue outliers have high Eddington ratios, high black hole masses, etc. ?),
and (3) for trends across our whole sample of BLS1 and NLS1 galaxies.  

\subsubsection{A strong correlation of line shift with ionization potential}

We detect a strong correlation between emission-line blueshift
and ionization potential (Fig. \ref{ionpot}). While the low-ionization
forbidden lines of [OI], [OII] and [NII], and the Balmer lines H$\alpha$
and H$\beta$ are at velocities very similar to [SII], high-ionization
lines on the other hand show strong blueshifts which increase with ionization potential. 
Two galaxies (SDSSJ115533.50+010730.4 and PG1244+026) show extreme blueshifts in [Fe X],
on the order of 600-1000 km\,s$^{-1}$. 
There is also a correlation between outflow
velocity and  
critical density of the individual line transitions (Fig. \ref{ionpot}), 
but it is less tight
in the sense that [OI]6300, which
has low ionization potential but high critical density, does not follow the trend.

\subsubsection{Trends and correlations of [OIII] blueshift with line and galaxy
parameters}

We have correlated the [OIII] blueshifts (of the blue outliers,
and of our sample as a whole) with various other line
parameters (line widths, line ratios) and with galaxy properties 
(absolute magnitude, black hole mass, Eddington ratio, NLR density).  
The strongest correlation is the one between [OIII] blueshift
and [OIII] line width (Spearman rank correlation coefficient
$r_{\rm S}$=0.6 for the whole NLS1 sample and even higher
for just the blue outliers; Fig. \ref{ionpot}). 
Within the NLS1 sample, there is a trend that blue outliers preferentially 
avoid low Eddington
ratios $L/L_{\rm edd}$ and low FeII/H$\beta$ ratios.  Most blue
outliers have small ratios of [OIII]/H$\beta_{\rm totl}$.   
We do not find trends of outflow velocity with 
NLR density $n_{\rm e}$, black hole mass, absolute
magnitude $M_{\rm i}$, and the width of the broad component of H$\beta$ (Fig. \ref{divcorr}).  
However, while trends are absent within the NLS1 population itself,
it is interesting to note that the presence of blue outliers
amplifies trends which become apparent when the BLS1 galaxies
are added to the correlation plots (Fig. \ref{divcorr}).  

\subsection{Frequency of blue outliers}
Among our sample, blue outliers only occur in NLS1 galaxies. 
At a velocity $v_{\rm [OIII]} \approxgt 150$ km\,s$^{-1}$,
we have a blue outlier fraction of 16\%. If we use $v_{\rm [OIII]} \approxgt 250$ km\,s$^{-1}$
instead, as in Zamanov et al. (2002), the fraction of blue outliers among NLS1
galaxies is 5\%. 

\section{Discussion}

\subsection{The locus of Seyfert galaxies with [OIII] blueshifts on the $M_{\rm BH}-\sigma$ plane}

Investigating the location of different types of galaxies on the $M_{\rm BH}-\sigma$ 
plane, and their potential evolution across the plane, 
is of great interest in the context of galaxy formation and evolution models.
In AGN, stellar velocity dispersion is often difficult to measure, and the width of
[OIII] has become a convenient proxy for stellar velocity dispersion
(e.g., Terlevich et al. 1990, Whittle 1992, Nelson \& Whittle 1996, Nelson 2000, Shields et al. 2003,
Boroson 2003, Greene \& Ho 2005, Netzer and Trakhtenbrot 2007, Salviander et al. 2007) after
removing [OIII] blue wings and excluding galaxies with powerful 
kiloparsec-scale linear radio sources. The scatter
in the relation is larger than in the original $M_{\rm BH}-\sigma_*$ relation
(Ferrarese \& Merritt 2000, Gebhardt et al. 2000), and indicates
secondary influences on the gas kinematics (e.g., Nelson \& Whittle 1996, Rice et al. 2006). 

Mathur et al. (2001) pointed out the importance of studying the locus
of NLS1 galaxies on the $M_{\rm BH}-\sigma$ plane, which was the focus
of a number of subsequent studies (Wang \& Lu 2001, Wandel 2002, 
Botte et al. 2004, 2005, Bian et al. 2004, Grupe \& Mathur 2004, 
Barth et al. 2005, Greene \& Ho 2005, Mathur \& Grupe
2005a,b, Zhou et al. 2006, Ryan et al. 2007, Watson et al. 2007, Komossa \& Xu 2007).  
Any such study would involve one important step:
the distinction between true outliers from the $M_{\rm BH}-\sigma$ relation on the one hand,
and apparent outliers on the other hand. Apparent outliers would only appear
to be off-set from the $M_{\rm BH}-\sigma$ relation because either the choice of 
line width as a measure of stellar velocity dispersion was
unsuitable, or the choice of line and continuum parameters
as a measure of black hole mass was unsuitable.   
KX07 have shown that those NLS1 galaxies of their sample
which deviate significantly from the $M_{\rm BH}-\sigma$ relation
of normal (and BLS1) galaxies are all characterized by high [OIII] blueshifts. 
While these [OIII] lines were therefore not suited as surrogate for stellar
velocity dispersion,  the [SII]-based measurements 
of velocity dispersion of the very same objects still located 
them on the $M_{\rm BH}-\sigma$ relation.    
Almost all galaxies which deviate most strongly from 
the $M_{\rm BH}-\sigma$ relation (see our Fig. 1) show $v_{\rm[OIII]} > 150$ km\,s$^{-1}$.
These six objects, plus three additional ones which also show $v_{\rm[OIII]} > 150$ km\,s$^{-1}$ 
define the nine blue outliers that are the target of the present study. 

In order to measure systematically the deviation of a galaxy from
the $M_{\rm BH}-\sigma_*$ relation, we define the quantity
$\Delta\sigma:= \log \sigma_{obs} - \log \sigma_{pred}$ (as in KX07),
where $\sigma_{obs}$ is the observed emission-line velocity dispersion,
and $\sigma_{pred}$ is the stellar velocity dispersion predicted from the $M_{\rm BH}-\sigma_*$ relation
of Ferrarese \& Ford (2005).    
We find a strong correlation between [OIII] blueshift $v_{\rm[OIII]}$ and $\Delta\sigma$
(Fig. 1).  
This finding demonstrates that [OIII] velocity
shift systematically affects the deviation of an object from 
the $M_{\rm BH}-\sigma$ relation (see also Boroson 2005). 
We further find that a correlation between $L/L_{\rm Edd}$ and $\Delta\sigma$ (e.g., Mathur \& Grupe 2005,
Greene \& Ho 2005; see also Netzer \& Trakhtenbrot 2007) is only present in our sample if we include blue outliers;
when they are removed from the sample, no correlation remains (last panel of our Fig. \ref{divcorr}, see also KX07).

Two of our NLS1 blue outliers have independent BH mass estimates. From X-ray variability, Czerny et al. (2001)
estimated $\log M_{\rm{BH}}=5.9$ for PG1244+062 which agrees well with our estimate
from applying the $R_{\rm BLR}$-$L$ relation ($\log M_{\rm{BH}}=6.2$).  
For RX01354$-$0043 we directly measured $\sigma_*$ from stellar absorption lines
(see Appendix). The value agrees well with $\sigma_{\rm [SII]}$ and puts
RXJ01354$-$0043 almost perfectly on the $M_{\rm{BH}}$-$\sigma_*$ relation (Fig. 1). 

These findings demonstrate the importance of measuring [OIII] blueshifts, and removing
objects with high blueshifts from a sample before putting it on
the $M_{\rm BH}-\sigma_{\rm [OIII]}$ relation. 
The independent question arises as to what may cause 
the blueshifts of these [OIII] outliers.

\subsection{Blue outliers: trends and correlations}

The phenomenon of slightly different blueshifts and
line widths of NLR emission lines has been recognized since the early
days of AGN spectroscopy and is generally
traced back to a certain stratification of the NLR
in the sense that high-ionization lines are produced preferentially
at small distances from the core, while low-ionization lines
are preferentially produced at larger radii (e.g., Osterbrock 1991).  
The presence of a strong [OIII] blueshift
on the order of several hundred km\,s$^{-1}$
was noticed early in 
the prototype NLS1 galaxy IZw1{\footnote{It is also 
known in some radio galaxies [e.g., PKS1549-79 (Tadhunter et al. 2001, Holt et al. 2006), 
IC5063 (Morganti et al. 2007), and 
PKS0736+01 (Zamanov et al. 2002, Marziani et al. 2003)].
Also note systematic blueshifts in high-ionization BLR emission lines
in quasar samples (e.g., Gaskell 1982, Sulentic et al. 2007).}}
(e.g., Phillips 1976,
V{\'e}ron-Cetty et al. 2004). The galaxy also shows very 
blueshifted IR coronal lines (Schinnerer et al. 1998) and blueshifted 
UV high-ionization broad lines (Laor et al. 1997).

Several recent studies systematically examined the phenomenon of blue outliers
in larger samples of type 1 Seyfert galaxies 
(Zamanov et al. 2002, Marziani et al. 2003, Aoki et al. 2005,
Boroson 2005, Bian et al. 2005). 
According to these studies blue outliers
have high Eddington ratios and small BLR Balmer line widths. 
However, not all sources
with high Eddington ratios are blue outliers. 
While Bian et al. (2005) further reported a
correlation between blueshift and Eddington ratio
of their 7 blue outliers, Aoki et al. (2005) did not 
find such a correlation for 16 objects.  
A strong correlation between
[OIII] blueshift and [OIII] line width is often seen, and is 
usually interpreted
as evidence for outflows. 
Zamanov et al. (2002) noticed that most of their blue outliers
also show high blueshifts in CIV1549 (but this phenomenon is not exclusive
to blue outliers; Sulentic et al. 2007).
Aoki et al. (2005) reported a trend that blue outliers
have high black hole masses ($> 10^7$ M$_{\odot}$). 
All previous 
studies of samples of blue outliers focused 
on the [OIII]-H$\beta$ spectral region  
in order to explore
the phenomenon, with the exception of 
Boroson (2005). Although Boroson (2005) measured 
the positions of other NLR lines in order to establish
a systemic reference for the [OIII] line properties, 
we have, for the first time, measured the widths 
and strengths of all the optical NLR lines in blue outliers
in order to study the nature of blue outliers and explore the dynamics
of the NLR.    
 
For each galaxy, we find a strong correlation between emission line blueshift
and ionization potential of the line-emitting ion. 
Coincidentally, higher ionization potentials of the ions in question also
come with higher critical densities of the forbidden line transitions
observed from the respective ions. Therefore, a correlation between outflow
velocity and ionization potential would generally
imply a correlation with critical density, and vice versa, raising
the question which of the two is the fundamental correlation;
density stratification or ionization stratification.  An important exception
to the above rule is [OI]6300, which has zero ionization potential, 
while its critical density  
($n_{\rm crit}=1.8\,10^{6}$ cm$^{-3}$) is relatively high.
[OI] is detected in several of our NLS1 galaxies, and we find that
[OI] follows the trend in ionization potential,
but not in critical density, arguing that the former correlation is the
underlying one (Fig. \ref{ionpot}){\footnote{It is well possible that we
have both, ionization and density stratification. In that case, the 
inner high-density
clouds are also highly ionized and/or matter-bounded, with little [OI] left.}}.
We confirm the previously known correlation between [OIII] blueshift and line width.

Blue outliers only occur among the NLS1
galaxies of our sample (i.e., sources with high Eddington 
ratios and small Balmer line widths),
but within the NLS1 population, blue outliers do not show a strong
correlation with Eddington ratio even though they preferentially
avoid low ratios.  
No  correlation of outflow velocity with black hole mass, absolute
magnitude, and with the width of the broad component of H$\beta$ is found.
Like in other samples,
the number of objects is small, and larger samples of blue outliers are needed
to confirm the weak trends.

It has occasionally been suggested that [OIII] in blue outliers appears broadened
by orientation effects in the sense that we look face-on on the central engine, and
into an outflow (Zamanov et al. 2002, Marziani et al. 2003, Boroson 2005).
If that scenario is correct, we can use blue outliers as test-beds for
orientation-dependent models of NLS1 galaxies which allow for the possibility
that the H$\beta$ line of NLS1 galaxies is narrowed due to viewing angle
effects. 
If, in blue outliers, the BLR was in a plane and if we viewed it face-on,
blue outliers should have the smallest H$\beta$ widths among NLS1 galaxies
(here, we temporarily classify NLS1 galaxies independently of their H$\beta$ width,
but only use the ratios [OIII]/H$\beta$ and FeII/H$\beta$ which still makes all
of our blue outliers NLS1 galaxies: in all cases [OIII]/H$\beta_{\rm totl} < 3$, and 
FeII/H$\beta_{\rm totl} > 0.5$). However, we do not find any trend
for small widths of the broad Balmer lines among the objects of our sample (see Fig. \ref{divcorr}).

The [OIII] lines of several objects still show the phenomenon of blue wings, arguing
against the previous suggestion that the classical [OIII] emission
in blue outliers is absent and we actually only see the blue wing.
[However, occasionally, the presence of a strong blue
wing could mimic a blue outlier, if the wing is
not spectroscopically resolved from the core of the line
(e.g., Grandi 1977, Holt et al. 2003)]. 

\subsection{Models of the narrow-line region}

There is good evidence that the motion of NLR clouds is
strongly influenced by 
the bulge gravitational potential 
(e.g., V{\'e}ron 1981, Whittle et al. 1992, Nelson \& Whittle 1996; see the prev. Sect. 4.1).
At the same time, there is also evidence for radial motions in NLRs
based for instance on blue asymmetries of [OIII] profiles
(e.g., de Robertis \& Osterbrock 1984, Veilleux 1991, Whittle 1992, V{\'e}ron-Cetty et al. 2001),
on correlations of blue wing blueshift with the Eddington ratio
and other arguments (Xu et al. 2007), and on spatially
resolved [OIII] velocity
shifts (e.g., Schulz 1990, Das et al. 2005). 
The existence of [OIII] blue wings 
can formally be equally well interpreted in terms of inflows
or outflows, plus selective obscuration. Generally, there is a preference for
the outflow interpretation.  

Regarding theoretical models for outflows, a lot of work has
focussed on magnetocentrifugal winds and radiatively driven
outflows from the accretion disk region (see K{\"o}nigl 2006, Everett 2007, Proga 2007
for reviews). 
These models have successfully been applied to BAL flows
and the BLR emission lines, are supported by recent observations
(Young et al. 2007), and have recently been extended to spatial
scales typical of the very inner NLR (Proga et al. 2008). However, little is known about 
the formation or continuation of such winds on much larger scales
on the order of 100 pc -- kpc characteristic for NLRs.    
Mechanisms suggested to be relevant in the NLR (and CLR) include 
radiation pressure acting on gas (Binette et al. 1997, Das et al. 2007) 
and on dust grains embedded in the 
clouds (Binette 1998, Dopita et al. 2002), 
and the entrainment of NLR clouds in hot winds
(Krolik \& Vrtilek 1984, Schiano 1986, Mathews \& Veilleux 1989, 
Smith 1993, Everett \& Murray 2007). 
The presence of
collimated outflows of radio plasma in form of jets
and their local interaction with NLR clouds
has been observed directly (e.g., Falcke et al. 1998).
Spatially resolved imaging spectroscopy
of nearby AGN 
has produced several good examples of spatial coincidences
between radio jets and line blueshifts (e.g., Riffel et al. 2006,
Morganti et al. 2007), while in other cases
the jet-cloud interaction
is very weak and does not significantly 
affect the local NLR velocity field (e.g., Cecil et al. 2002, Das et al. 2005). 
In some cases, jets are absent but emission lines are still blueshifted
(Barbosa et al. 2008, in prep).  
In AGN with strong starburst
activity winds could also be starburst driven (e.g., Rupke et al. 2005).

Regarding NLS1 galaxies, their high Eddington ratios
make the presence of radiation-pressure driven {\em winds} (on the accretion disk scale) 
very likely. Less certain and understood is the efficiency of {\em jet} launching
under high-accretion rate conditions. Again, these mechanisms refer to the innermost AGN region,
not to larger scales.  
Blue outliers with their extreme velocity shifts place
particularly tight constraints on models for AGN outflows on large, i.e. NLR, scales.  
We discuss several NLR models for blue outliers in turn. 

\subsubsection{A compact NLR ?} 
Zamanov et al. (2002) and Marziani et al. (2003) suggested that blue outliers
posses a very compact NLR, perhaps the result of the youth of these NLS1
galaxies which did not yet develop a full NLR. 
Their idea was based on the assumption that blue outliers would only posses [OIII] emission
and lack other narrow emission lines.   
However, we do find
a classical NLR in terms of the presence of low-ionization lines.
Further, the fact that their [SII] line widths put the blue outliers on the same $M_{\rm BH}-\sigma$ relation
with BLS1 and normal galaxies strongly indicates that their outer NLR is a quiescent
classical NLR similar to that of other BLS1 and NLS1 galaxies.  

\subsubsection{A two-component NLR} 
The strong blueshifts 
we detect in the high-ionization component raises
the possibility that we see two independent components;
a classical NLR, plus an independent outflow component
which could be due to a disk wind,
a wind from the torus 
or in form of a lowly ionized warm absorber.   
However, 
this interpretation is very unlikely, because the non-blueshifted [OIII] counterpart 
to the low-ionization lines, as it would be expected from 
{\em the inner part} of a classical NLR,
is weak or absent. This fact leads us almost 
inevitably to the third and fourth possibility. 

\subsubsection{A classical NLR, modified at small radii}
The fact that we do not detect a classical inner NLR, in form of a
second, non-blueshifted
[OIII] emission component, implies that processes directly
modify/disturb at least the inner NLR; perhaps in form of
radio-jet -- cloud interaction.  
However, 
the strong dependence of emission line velocity shift on ionization
potential that we detect (Fig. \ref{ionpot}) indicates that we see a highly stratified, photoionized
medium, not a local interaction due to shocks and locally disturbed
velocity fields.  

\subsubsection{NLR clouds entrained in a wind} 
The phenomenologically most straight-forward solution 
is one in which the (CLR and) NLR clouds themselves follow a decelerated
outflow. This scenario requires an efficient driving mechanism
and perhaps an efficient deceleration mechanism.  
One possibility is direct acceleration of the NLR clouds 
by radiation pressure acting on gas or dust; another is 
the presence of a hot wind 
which entrains the NLR clouds (see Sect. 4.4).  
If the wind forms
a decelerating outflow, the entrained clouds of the  
CLR/ inner NLR
would have the highest outflow velocities, respectively, which would decrease
in dependence of core distance 
and would leave the outer, low-ionization part of the NLR unaffected. 

Zamanov et al. (2002) reported a linkage of [OIII] blue-outlierness 
with broad-line CIV
blueshift. Could the same wind persist from the disk,
to
the high-ionization
BLR [but leave the bulk of the BLR
unaffected; by blowing perpendicular ?], and on to the CLR and inner NLR ?

In this picture, the NLRs of BLS1 galaxies, NLS1s, and blue outliers would be
intrinsically similar, and the motion of NLR clouds generally dominated
by the bulge gravitational potential. 
Which emission line regions partake in an outflow would depend on the 
efficiency/operating distance 
of the wind. It could be present in all AGN, but would generally only affect the CLR (in AGN
with blueshifted iron coronal lines but quiescent NLRs),
or level off even before reaching the CLR. 

\subsection{Mechanisms of cloud acceleration and entrainment}

While phenomenologically successful, the question is raised as to the
origin of the wind/outflow on the one hand, and the entrainment mechanism
and cloud stability against disruptive instabilities (Mathews \& Veilleux 1989,
Schiano et al. 1995) 
on the other hand.
We discuss several possibilities in turn. 

\paragraph{Cloud acceleration by radiation pressure acting on dust}
Radiation pressure acting on dust grains embedded in the gas clouds
is an efficient way of cloud acceleration (e.g., Binette 1998,
Dopita et al. 2002, Fabian et al. 2006). 
In blue outliers, this mechanism is likely not at work, because
we do not expect dust in the high-ionization BLR (we are assuming
here that the high-ionization part of the BLR partakes in the outflow, motivated
by the results of Zamanov et al. (2002) described in Sect. 4.3.4).
Some dust might be present in the CLR (even a small admixture 
of dust would still lead to efficient radiative acceleration,
and generally still predict sufficiently strong gas-phase iron lines; Binette 1998). 
While, observationally, dust-rich NLRs could be present in individual blue outliers
(e.g., RXJ0135$-$0043), most of them have very blue optical continua
arguing against the presence of an excess of dusty gas along the line of sight. 
These arguments make cloud acceleration by radiation pressure acting
on dust grains an unlikely scenario. 

\paragraph{Cloud entrainment in collimated radio plasma}

Outflowing radio plasma in form of jets is known
to be present and to reach large
distances from the nucleus in AGN, so could plausibly affect
several emission-line regions on its way.  
The phenomenon of cloud entrainment in radio plasma has been directly
observed and studied 
in starforming regions (Ostriker et al. 2001, Stojimirovi{\`c} et al. 2006).
Jet-cloud interaction can result in a variety of phenomena, including
cloud entrainment, jet deflection and disruption and 
cloud destruction (e.g., Saxton et al. 2005, Krause 2007 
and references therein). 
Of interest here is entrainment (Blandford \& K{\"o}nigl 1979,
Schiano et al. 1995, Fedorenko et al. 1996). One key problem is cloud
longevity against various instabilities.  
Fedorenko et al. (1996) argued for magnetic NLR confinement 
which would then allow for NLR cloud entrainment in radio jets. 

Independent of theoretical considerations the question is
raised if the blue outliers of our sample all harbor powerful radio jets. 
Relatively little is known about the radio
properties of NLS1 galaxies in general. On average, they
tend to be less radio-loud than BLS1 galaxies (Zhou et al. 2006,
Komossa et al. 2006)
and share some similarities with compact steep spectrum radio sources
(Komossa et al. 2006).
Four of the blue outliers of our sample have FIRST radio detections.
These imply radio powers
of $P_{\rm 1.4}$= 10$^{22}$--8\,10$^{23}$ W\,Hz$^{-1}$ (Tab. 1) at 1.4 GHz; similar
to those AGN which Nelson \& Whittle (1996) find to be off-set from
the $M_{\rm BH}-\sigma_*$ relation. We do not have information on the radio
morphology of the blue outliers. Sources are unresolved with FIRST,
with one remarkable exception: the radio emission of RXJ0135$-$0043
is extended by $\sim$10 kpc (or double; see Appendix).
Spatially resolved radio observations of the sources are required to
search for the presence of jets and to study the radio
properties in more detail.

\paragraph{Thermal winds} 
Variants of thermal wind models have been studied
in order to explain the kinematics of ionized absorbers
(Chelouche \& Netzer 2005), and of NLRs. 
{\em Isothermal} Parker wind models of Everett \& Murray (2007),
originally computed in order to model spatially resolved
NLR velocity gradients of NGC\,4151 (Das et al. 2005), would
have roughly the right properties to explain our average blue outlier
velocity shifts and radial velocity changes.  
A range in NLR cloud column densities would lead to
a spread in cloud velocities (see Eqn. (10) of Everett \& Murray 2007),
thereby perhaps explaining the observed line broadening. 
However, as also shown by Everett \& Murray (2007), realistic models
including photoionization heating have temperature gradients, 
and adiabatic cooling decelerates the winds too  quickly.  
An extra heating source of unknown nature would be needed 
in order to keep the wind isothermal. 
Models of Das et al. (2007), based
on radiative acceleration of NLR clouds, implied that 
NLR clouds do not decelerate quickly enough 
in order to explain NLR velocity gradients of NGC\,1068 observed with HST;
models work after introducing drag forces from an ambient medium.  
Detailed modeling of the present data would likely involve
several of the above model ingredients.
Such modelling is beyond the scope
of this paper.   

\paragraph{High Eddington ratios and orientation effects}
Common to blue outliers is their high Eddington ratios.
Could high $L/L_{\rm Edd}$ be the wind 
driving mechanism ?  
Additional orientation
effects (as also discussed by Marziani et al. 2003, Boroson 2005) 
are still needed in order to explain why not all AGN with high $L/L_{\rm Edd}$
are blue outliers. 
Near face-on orientation would have us look more down the flow in blue outliers, 
thereby enhancing the stratification, broadening, and blueshift effect.

\subsection{Links with (compact) radio sources and mergers}

It is interesting to point out similarities between
the phenomenon of blue outliers
in NLS1 galaxies and in radio galaxies (see also Holt et al. 2006).
Several radio galaxies  show a similar phenomenon of
high [OIII] core blueshifts and line broadening (e.g., Tadhunter et al. 2001,
Marziani et al. 2003, Holt et al. 2006, Stockton et al. 2007;
see Gupta et al. 2005 for a related phenomenon
in the UV). 
Jet-cloud interaction in the (inner) NLR
is the favored interpretation of most of these sources.  
These objects may represent the early stages of radio-source evolution
(Tadhunter et al. 2001). 
Radio galaxies with blue outliers in [OIII]
are typically compact flat spectrum sources, are absorbed,  
and are luminous in the infrared.  For some (e.g., PKS1549$-$79 and
PKS0736+01)
there is evidence that we have a near face-on view (Tadhunter et al. 2001, Marziani et al. 2003).
As pointed out before, the strong stratification we see in NLS1 blue outliers
argues against local jet-cloud interactions in those galaxies, but would be
consistent with NLR clouds entrained in the outflowing radio plasma.
 
The blue outliers among the radio galaxies show signs
of recent mergers.  Regarding NLS1 galaxies in general, there is no
evidence that the majority of them underwent recent mergers or has an excess of 
companion galaxies (Krongold et al. 2001, Ryan et al. 2007).
Little is known about the host galaxies of our blue outlier sample, in particular.
IRAS11598-0112
indeed is a merger with prominent tidal tails (Veilleux et al. 2002),
ultraluminous in the infrared.  Inspecting the SDSS images
of the other galaxies of our sample we do not find strongly
disturbed galaxy images indicating ongoing mergers, but this
needs to be confirmed with deeper imaging.  One galaxy, RXJ0135$-$0043,
shows indications of an off-center nucleus. That effect could
be caused by interaction, be mimicked by dust, or have another origin.   

\subsection{Recoiled Black Holes ?}
Recent simulations of merging black holes predict black hole recoil velocities
due to emission of gravitational wave radiation 
up to several thousand km\,s$^{-1}$ (e.g., Campanelli et al. 2007; review by Pretorius 2007). 
Potentially, high relative outflow velocities of AGN emission
line regions vs the host galaxy can arise if the recoiled
black hole keeps its BLR and inner NLR (Bonning et al. 2007). 
Applied to blue outliers, if they harbored recoiled BHs,
their whole BLR (plus the high-ionization NLR) 
should show high blueshifts, 
while the `remnant' NLR would still appear
in low-ionization lines.   In that case, we expect the broad component
of the Balmer lines to exhibit the highest blueshifts of all emission lines;
which is, however, not observed. Shifts in broad H$\beta$ are
less than those in [OIII].

\subsection{Future work}

The SDSS data base is well suited for a systematic search
for more blue outliers. Zhou et al. (2006) mention in passing
the presence of several extreme ones among their NLS1 galaxy sample but
do not discuss them further.  
Spatially resolved optical spectroscopy
will allow us to measure directly  
line widths, outflow velocities, etc. in dependence of 
the core distance{\footnote{This approach is difficult, however,
if the near-pole-on interpretation of blue outliers is correct.}}.  Spectroscopy in the IR and UV 
will tell whether trends (a correlation with ionization potential) 
persist in IR coronal lines
and high-ionization UV broad lines.    
X-ray measurements of blue outliers will be useful to search for
signs of high-velocity ionized outflows,
and to see whether blue outliers
again stick out in AGN correlation space when adding their 
X-ray properties (X-ray steepness, variability) to
correlation analyses. 
(High-resolution) radio observations of all galaxies
will facilitate further comparison between blue outliers
in radio galaxies and NLS1 galaxies and will tell whether radio jets
are present in NLS1 blue outliers. Radio observations 
have the potential to confirm the pole-on hypothesis of
blue outliers, if relativistic beaming is detected.  
Imaging with HST will reveal  
whether the host galaxies of blue outliers show signs of recent
mergers.  
In particular, galaxy merger simulations predict strong outflows in the
final merger phase (e.g., Springel et al. 2005). Imaging will allow us to
test whether blue outliers are in such a phase. 

If the face-on interpretation of blue outliers is correct,
they are also useful test-beds for the question, whether the width
of broad H$\beta$ is systematically affected by orientation. 
Increasing the sample size will allow us to test more stringently
whether the broad component of H$\beta$ is systematically narrower
in blue outliers. 
If so, this would imply that BLR clouds are
arranged in a plane, 
and we would underestimate systematically the BH masses in these objects,
and perhaps NLS1 galaxies in general.
If, instead, their BLR is spherical, we would not see systematically
narrower H$\beta$ in objects viewed face-on, and would not have
to worry about correctness of BH mass estimates.   Our preliminary
results indicate that the latter is the case. 

On the theoretical side, the question is raised as to which
winds can operate across long distances spanning the high-ionization BLR,
CLR and a substantial part of the NLR,
predict the observed gradient in cloud velocities 
($\sim$1000 km\,s$^{-1}$ of high-ionization emission-line clouds
close to the nucleus, and several hundred km\,s$^{-1}$ further out
on typical NLR scales while the outer NLR is mostly unaffected), and ensure
longevity of the emission-line clouds.

\section{Summary and conclusions}

In [OIII] blue outliers  with their extreme velocity shifts 
the effects of secondary influences on the NLR kinematics
are enhanced or dominate completely. 
Their study is therefore of great relevance for 
(1) scrutinizing the usefulness and limitations of [OIII] width as a proxy 
for stellar velocity dispersion; (2) understanding the origin
and dynamics of the NLR; (3) investigating driving mechanisms 
of AGN outflows on large scales; and (4) examining possible links with 
results from merger simulations which predict that a substantial fraction
of the ISM of
the merger should be outflowing.  
We have systematically studied the optical properties of
such AGN with high [OIII] blueshifts which deviate from the 
$M_{\rm BH}-\sigma_{\rm [OIII]}$ relation of BLS1 and NLS1 galaxies,
and obtained the following results:  

$\bullet$ All of them have high Eddington ratios ($L/L_{\rm{edd}}$=0.5--1.5) and narrow BLR Balmer lines 
(FWHM(H$\beta_{\rm b}$)=1200--1800 km\,s$^{-1}$),
which makes them NLS1 galaxies. 
The fraction of blue outliers among our NLS1 sample 
is 16\% ($v_{\rm [OIII]} \approxgt 150$ km\,s$^{-1}$), 
and 5\% at the highest outflow velocities ($v_{\rm [OIII]} \approxgt 250$ km\,s$^{-1}$). 
While blue outliers 
do enhance correlations which appear across
the whole BLS1-NLS1 population,  
we do not find
strong correlations of [OIII] outflow velocity with the Eddington ratio
within the blue outlier population itself; perhaps due to the small sample size.   

$\bullet$ We do detect a strong correlation between emission-line blueshift and ionization
potential, and confirm a strong correlation between 
[OIII] blueshift and [OIII] line width.  
The presence of a classical quiescent {\em outer} NLR  
is indicated by the existence of low ionization lines, by
[SII] line widths which locate the blue outliers on the 
same $M_{\rm BH}-\sigma_{\rm [SII]}$ relation as other
BLS1 and NLS1 galaxies, and by inferred NLR densities similar
to other NLS1 galaxies.  
On the other hand, zero-blueshift [OIII] emission expected from a quiescent 
{\em inner} NLR is weak or absent.  

$\bullet$ Taken together, these observations place tight constraints on
models: We favor a scenario where NLR clouds of blue outliers 
are entrained in a decelerating wind. 
Similar, less powerful winds
could be present in all AGN, but would generally only affect the CLR (in AGN
with blueshifted iron coronal lines but quiescent NLRs),
or level off even before reaching the CLR only affecting the high-ionization BLR.

$\bullet$ The mechanism that drives and decelerates the wind is speculative at
present, but could be linked to the high Eddington ratios of the galaxies. 
Extra orientation effects (near pole-on views), 
considered previously to explain the
correlation of [OIII] blueshift with line width of blue outliers, 
would also explain the strong ionization stratification we detect. 

$\bullet$ Two blue outliers have independent BH mass / stellar velocity
dispersion measurements and these place them on or close to the 
$M_{\rm BH}-\sigma$ relation of non-active galaxies. 
This, together with the fact that the width of broad H$\beta$
does not correlate with [OIII] outflow velocity, indicates:
If blue outliers are indeed seen more face-on, this fact does
not reflect strongly in their H$\beta$ widths, implying 
that their BLR geometry is closer to spherical than to planar.

$\bullet$ Most remarkable among the blue outliers is the galaxy RXJ01354$-$0043.
Unlike other NLS1 galaxies its radio emission is extended and possibly
double, its optical Balmer lines appear to be double-peaked, and its optical
spectrum shows strong absorption lines from the host galaxy.
The link between blue outliers in NLS1
galaxies and in (compact) radio galaxies needs further exploration.

\section{Appendix: Notes on individual objects}

A few sources are included in the SDSS-NLS1 samples of
Williams et al. (2002) and Anderson et al. (2003).  
The high [OIII] blueshifts of SDSSJ115533.50+010730.4
and RXJ01354$-$0043 were reported by Bian et al. (2005). 
Boroson (2005) measured [OIII] blueshifts of NGC450\#86, RXJ01354$-$0043, 
PG1244+026, and SDSS17184+5734. 
Here, we provide a short summary of the multi-wavelength
properties of the blue outliers collected from the literature,
and give some comments based on our optical spectral
analysis. Galaxies are listed in
order of decreasing [OIII] blueshift. 

\paragraph{SBS0919+515} This AGN is a known X-ray source,
first detected with the Einstein observatory (Chanan et al. 1981),
mostly studied in the X-ray regime (e.g., Boller et al. 1996,  
Vaughan et al. 2001), and optically identified as NLS1 galaxy by Stephens (1989). 
It shows the highest [OIII] blueshift of our sample ($v_{\rm [OIII]}=430$ km\,s$^{-1}$). 

\paragraph{SDSSJ115533.50+010730.4} This AGN was detected in X-rays during
the ROSAT all-sky survey (Voges et al. 1999) and was first classified as 
NLS1 galaxy by Williams et al. (2002).  
It has the second-highest [OIII] velocity shift of our sample
($v_{\rm [OIII]}=330$ km\,s$^{-1}$), highly blueshifted [NeV], 
and faint coronal line emission of [FeX] with a blueshift 
corresponding to $v_{\rm[FeX]} \approx 1000$ km\,s$^{-1}$,
among the highest blueshifts reported for coronal lines to date. 

\paragraph{RXJ01354$-$0043 (SDSSJ013521.68$-$004402.2)} This AGN was detected in the X-ray
and radio band (Brinkmann et al. 2000, Wadadekar 2004)
and was optically identified as NLS1 galaxy by
Williams et al. (2002). 
It is remarkable in several respects. 
We find that its optical spectrum 
appears to be dominated by the host galaxy. Strong absorption
lines from higher order Balmer lines,
Ca H\&K and NaID are detected (Fig. 5).  
The optical AGN continuum emission is either intrinsically weak, or absorbed (which
would require a peculiar geometry of the absorber, given that
the broad Balmer lines are present).   
RXJ01354$-$0043 is detected with GALEX{\footnote{http://galex.stsci.edu/GR2/}} and displays red UV colors. 
Its (narrow) Balmer lines appear double-peaked, with a separation
of $\sim$580 km\,s$^{-1}$,  
or else
are affected by residual, exceptionally strong
and redshifted, host-galaxy features which is very unlikely, given the strength
of H$\alpha$.
The line decomposition shown in Fig. 5 assumes
the same widths of the narrow core of H$\beta$ and of the two [NII] lines,
fixed to the width of [SII], and further assumes a fixed [NII] line ratio. 
The remaining profile was modeled with two Gaussians of free parameters
and can be well fit with a broad component plus a second, redshifted relatively
narrow component.
RXJ01354$-$0043 has a FIRST detection with a radio flux of 2 mJy 
and there is evidence for extended radio emission
(Becker et al. 1997{\footnote{most recent catalogue
update at http://cdsarc.u-strasbg.fr/viz-bin/Cat?VIII/71}}), corresponding to
a scale of $\sim$10 kpc. 
This is remarkable because few if any NLS1 galaxies are
known to have widely extended radio emission 
(Ulvestad et al 1995, Komossa et al. 2006, Yuan et al. 2008). 
Inspecting the FIRST image cut-out, the source appears
to be double.   
Among the emission lines, only the Balmer lines appear double-peaked. 
RXJ01354$-$0043 is   
the first blue outlier to show this phenomenon.
Explanations include jet-cloud interaction,  
a bipolar outflow or, speculatively, superposed strong Balmer absorption
(which is rarely seen because the required level population is not
easy to achieve; Hall 2007, Lu et al. in prep.).
We have used the stellar absorption lines from the host galaxy for a 
measurement of $\sigma_*$. We obtain $\sigma_*=82$ km\,s$^{-1}$    
which is in good agreement with the [SII]-based 
measurement, $\sigma_{\rm [SII]}=106$ km\,s$^{-1}$, 
and locates RXJ01354$-$0043
(almost perfectly) on the $M_{\rm BH}-\sigma_*$ relation of non-active galaxies (Fig. 1).   

\paragraph{NGC450\#86} Identified as NLS1 galaxy by Williams et al. (2002)
and
detected in X-rays during the ROSAT all-sky survey
(Voges et al. 1999). 

\paragraph{SDSSJ032606.75+011429.9} A basically unknown galaxy,  
with optical NLS1 spectrum (Williams et al. 2002). 
No radio or X-ray detection was reported.  

\paragraph{IRAS\,11598$-$0112} This galaxy is ultraluminous 
in the infrared (Murphy et al. 1996, Kim \& Sanders 1998)
with a single nucleus and prominent tidal tails (Veilleux et al. 2002).  
Its optical spectrum is that of a NLS1 galaxy (Moran et al. 1996).  
It is almost radio-loud with a 1.4 GHz radio index $R \approx 5$ (Komossa et al. 2006)
and was first detected in X-rays with ROSAT (Voges et al. 1999).
The FeII emission in its optical spectrum is not well described by our
FeII template. In particular, the `red' and `blue' complexes
of FeII do not match each other well, and the source redshift appears to differ
from that of FeII.
Strong emission of unknown
nature remains blueward of [OIII]. Formally, we can describe it with
an extra [OIII] component, broad (FWHM([OIII])=1800 km\,s$^{-1}$) and highly blueshifted ($v$=1300 km\,s$^{-1}$). 
If real, it could represent an
extreme blue wing, or could imply the presence of an extra starburst/shock-driven component.
The structure needs to be confirmed by
independent spectroscopy.

\paragraph{SDSSJ171828.99+573422.3} Detected with ROSAT (Brinkmann et al. 1999),
and optically identified as NLS1 by Williams et al. (2002). 

\paragraph{PG1244+026} This AGN (Green et al. 1986) is a well-known 
X-ray, UV, infrared and radio source (e.g., Elvis et al. 1986, Kellerman et al. 1989,
Sanders et al. 1989, Fiore et al. 1998, Ballantyne et al. 2001, Jim{\`e}nez-Bail{\`o}n et al. 2005)
with an optical NLS1 spectrum (Miller et al. 1992, Veron-Cetty et al. 2001). 
It is radio quiet with a 5 GHz radio 
index of $R=0.5$ (Kellerman et al. 1989). Hayashida (2000) and Czerny et al. (2001)
determined its black hole mass based on X-ray variability (the power density spectrum).
Czerny et al. (2001) report $\log M_{\rm{BH}}=5.9$ 
which is consistent with our
value of $\log M_{\rm{BH}}=6.2$ from applying the relation of Kaspi et al. (2005). 
We detect in its optical spectrum [FeX] emission with an outflow velocity of 640 km\,s$^{-1}$.  

\paragraph{RXJ09132+3658} Detected in X-rays with ROSAT
(Voges et al. 1999, Brinkmann et al. 2000) and in the radio band during
the FIRST survey (Becker et al. 1997),  and
identified as NLS1 galaxy by Xu et al. (1999). 


\acknowledgments
DX acknowledges 
the support of the Chinese
National Science Foundation (NSFC)
under grant NSFC-10503005, and 
the support of MPG/MPE.
HZ acknowledges support from the Alexander von Humboldt Foundation,
from NSFC (grant NSF-10533050), and from program 973 (No. 2007CB815405).
LB acknowledges support from CONACyT grant J-50296.      
We thank our referee for his/her comments and suggestions,
and the members of MPE's new Physics of Galactic Nuclei group
and J. Sulentic, D. Proga and D. Merritt 
for discussions. 
This research has made use
of the SDSS data base, and of the NASA/IPAC Extragalactic Database (NED) 
which is operated by the Jet Propulsion Laboratory, California Institute of
Technology, under contract with the National Aeronautics and Space Administration. 
Funding for the SDSS and SDSS\,II has been provided by the Alfred P. Sloan Foundation, 
the Participating Institutions, the National Science Foundation,
the U.S. Department of Energy, the National Aeronautics and 
Space Administration, the Japanese Monbukagakusho, the Max Planck Society, and the Higher
Education Funding Council for England.
The SDSS is managed by the Astrophysical Research Consortium 
for the Participating Institutions. The Participating Institutions are the American
Museum of Natural History, Astrophysical Institute Potsdam, 
University of Basel, University of Cambridge, Case Western Reserve University, University
of Chicago, Drexel University, Fermilab, the Institute for Advanced Study, 
the Japan Participation Group, Johns Hopkins University, the Joint
Institute for Nuclear Astrophysics, the Kavli Institute for 
Particle Astrophysics and Cosmology, the Korean Scientist Group, the Chinese Academy of
Sciences (LAMOST), Los Alamos National Laboratory, the 
Max-Planck-Institute for Astronomy (MPIA), the Max-Planck-Institute for Astrophysics (MPA), New
Mexico State University, Ohio State University, University of Pittsburgh, 
University of Portsmouth, Princeton University, the United States Naval
Observatory, and the University of Washington.


\begin{deluxetable}{llcccccccccc}
\tabletypesize{\tiny}
\tablecaption{Properties of [OIII] blue outliers}
\tablewidth{0pt}
\tablehead{  
\colhead{coordinates\,(J2000)} & \colhead{common name} & \colhead{$z$} &  \colhead{$\Delta v$} & \colhead{w(OIII$_{\rm c}$)} & \colhead{w(H$\beta_{\rm b}$)} & \colhead{R5007} &\colhead{R4570} & \colhead{$M_{\rm BH}$} & \colhead{$L/L_{\rm Edd}$} & \colhead{$n_{\rm e}$} & \colhead{$P_{\rm 1.4}$} \\
\colhead{(1)\tablenotemark{a}} & \colhead{(2)} & \colhead{(3)} & \colhead{(4)} & \colhead{(5)} & \colhead{(6)} & \colhead{(7)}  & \colhead{(8)}  & \colhead{(9)}  & \colhead{(10)}  & \colhead{(11)} & \colhead{(12)}   
}
\startdata
092247.03 +512038.0 & SBS0919+515    & 0.161  &430 & 720 & 1250 & 0.3 & 1.3 & 6.7 & 1.5 & 10 & \\
115533.50 +010730.6 & SDSSJ11555+0107& 0.198 & 330 & 780 & 1510 & 0.3 & 0.7 & 6.7 & 0.9 & 200 & \\
013521.68 $-$004402.2 & RXJ01354$-$0043  & 0.099 & 240 & 620 & 1710 & 1.1 & 0.5 & 6.5 & 0.5 & 90 & 22.7 \\
011929.06 $-$000839.7 & NGC 450\#86    & 0.091 & 220 & 380 & 1220 & 0.5 & 0.9 & 6.2 & 0.9 & 200 &  \\
032606.75 +011429.9 & SDSSJ03261+0114& 0.128 & 180 & 530 & 1230 & 0.5 & 0.9 & 6.3 & 1.0 & 340 & \\
120226.76 $-$012915.3 & IRAS11598-0112 & 0.151 & 170 & 340 & 1460 & 0.5 & 2.7 & 6.8 & 1.1 & 110 & 22.9 \\
171829.01 +573422.4 & SDSSJ17184+5734& 0.101&  150 & 470 & 1760 & 0.4 & 0.7 & 6.6 & 0.5 & 30 & \\
124635.25 +022208.8 & PG1244+026     & 0.049 & 150 & 300 & 1200 & 0.6 & 0.8 & 6.2 & 0.9 & 400 & 22.1\\
091313.73 +365817.3 & RXJ09132+3658  & 0.108 & 150 & 350 & 1680 & 1.0 & 0.5 & 6.5 & 0.5 & 40 & 22.5 
\enddata
 \tablenotetext{a}{columns from left to right: (1) SDSS optical
coordinates in RA (h,m,s) and DEC (d,m,s), (2) galaxy name, (3) redshift determined from H$\beta$, 
(4) [OIII]$_{\rm core}$ velocity (blueshift) wrt. [SII] in km\,s$^{-1}$, (5) FWHM([OIII]$_{\rm c}$)
in km\,s$^{-1}$, (6) FWHM(H$\beta_{\rm b}$) in km\,s$^{-1}$,
(7) ratio of total [OIII] over total H$\beta$ emission, (8) ratio of FeII4570 over total H$\beta$ emission,
(9) logarithm of black hole mass in solar masses, (10) Eddington ratio, (11) NLR electron density in cm$^{-3}$,
(12) logarithm of the radio power at 1.4 GHz in W/Hz of galaxies detected in the FIRST survey. }

\end{deluxetable}


\begin{figure*}
\plotone{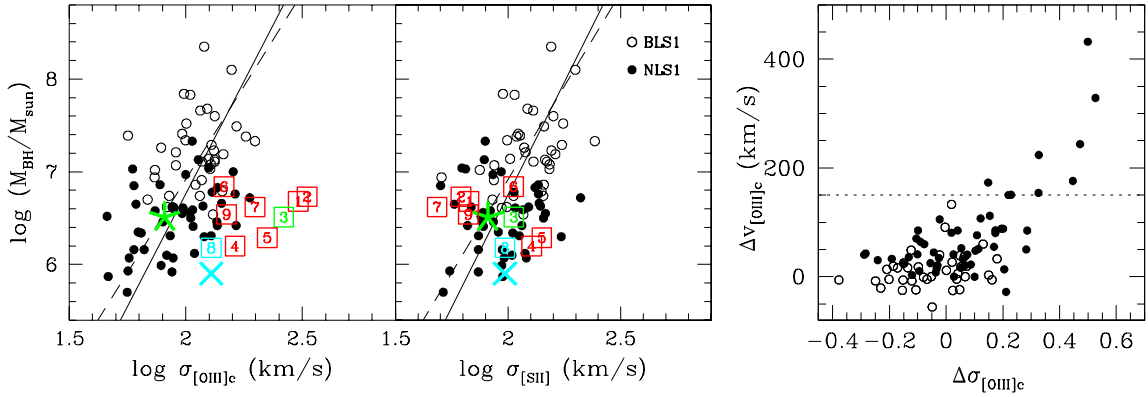} 
\caption{ 
Location of blue outliers (open squares) 
on the $M_{\rm BH}-\sigma$ plane
in comparison to the full NLS1 (filled circles)
and BLS1 (open circles) sample
(Komossa \& Xu 2007). 
{\sl{Left}}: The leftmost panel is based on $\sigma$ measurements from the narrow core of [OIII]$\lambda$5007
(asymmetric blue wings were removed). 
Blue outliers in [OIII] (with
radial velocities larger than 150 km\,s$^{-1}$) are marked with an open square. 
Object names are
coded by number (1=SBS0919+515, 2=SDSSJ11555+0107, 3=RXJ01354$-$0043, 4=NGC\,450\#86, 5=SDSSJ03261+0114,
6=IRAS11598$-$0112, 7=SDSSJ17184+5734, 8=PG1244+026, 9=RXJ09132+3658). 
The second panel shows the same relation based on [SII]. 
NLS1 and BLS1 galaxies follow the same $M_{\rm BH}-\sigma_{\rm{[SII]}}$ relation.
The dashed and solid lines represent the $M_{\rm BH}-\sigma_*$ relation of non-active
galaxies of Tremaine et al. (2002) and
of Ferrarese \& Ford (2005; FF05),
respectively. 
Two objects with independent BH mass (PG1244+026) and stellar velocity dispersion
(RXJ01354$-$0043) estimates are marked with special symbols (blue cross: PG1244+026,
green asterisk: RXJ01354$-$0043). 
{\sl{Right}}: Deviation $\Delta\sigma$ of NLS1 and BLS1 galaxies 
from the $M_{\rm BH}-\sigma_*$ relation of FF05 in dependence of [OIII] radial velocity, $\Delta$v,
measured w.r.t. [SII].  
}
\label{fig1}
\end{figure*}


\begin{figure*}
\plotone{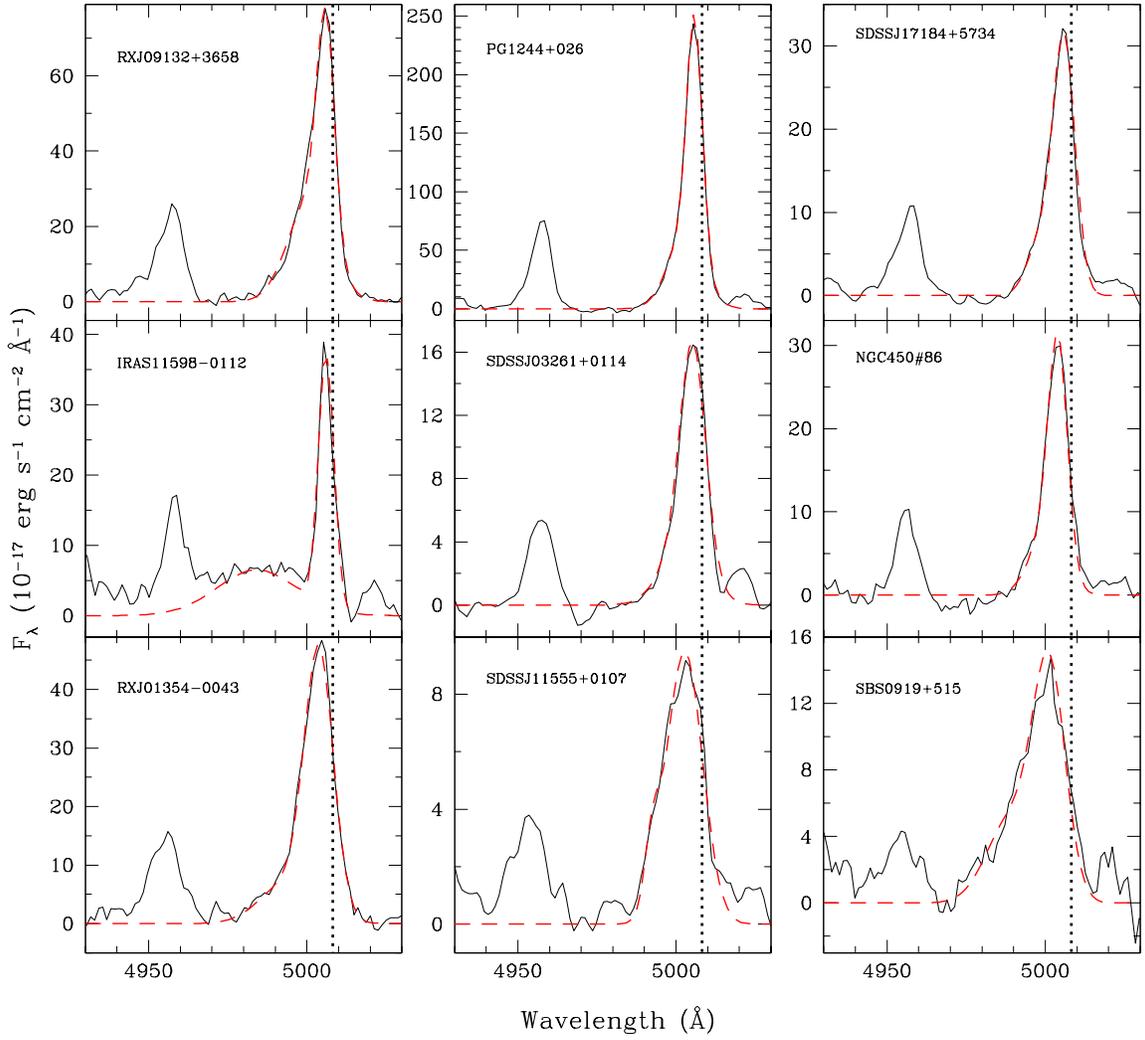}
\caption{[OIII]4959,5007 emission-line region of all nine blue outliers (FeII and continuum subtracted).
The expected [OIII] peak location according to [SII] is marked with the dotted line. 
Our two-component Gaussian fit to each [OIII]5007 emission line is overplotted using a dashed line.
}
\label{o3profiles}
\end{figure*}


\begin{figure*}
\plotone{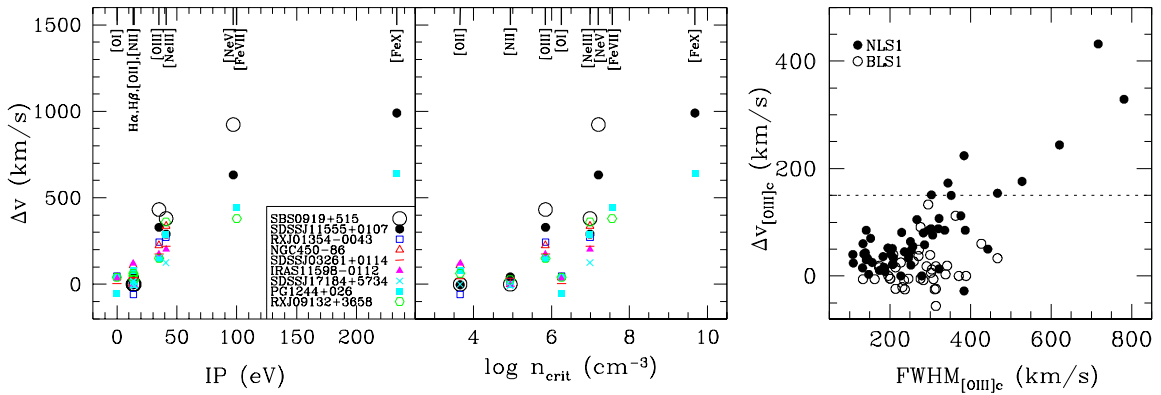} 
\caption{Velocity shift of individual emission
lines versus the ionization potential IP and the critical density $n_{\rm crit}$ (left), and correlation
of the [OIII] line width with the [OIII] blueshift (right). 
}
\label{ionpot}
\end{figure*}

\begin{figure*}
\plotone{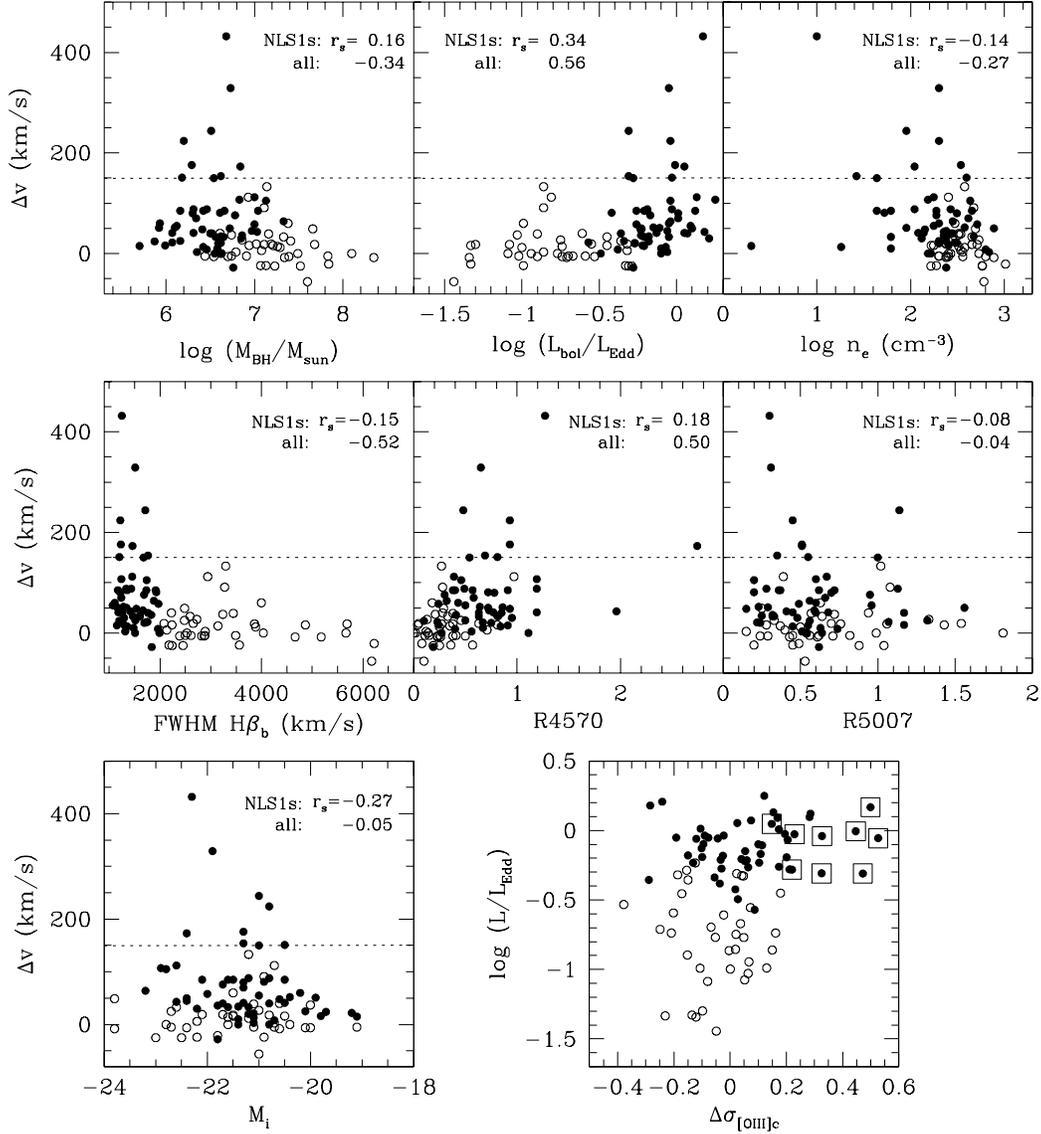} 
\caption{Correlation of [OIII] outflow velocity with other emission-line and AGN properties (panels 1-7).
Objects above the dashed line are blue outliers. NLS1 galaxies are represented by filled
circles, BLS1 galaxies by open circles. The correlation coefficient $r_{\rm s}$ shown in
each panel was calculated among the NLS1 population only, or for the whole
sample (marked as ``all:'' in the graphs). If the BLS1 galaxies are included,
several correlations emerge, and the blue outliers amplify these trends.  
R5007 corresponds to the ratio of total [OIII] over total H$\beta$ emission, R4570
to the ratio of FeII4570 over total H$\beta$ emission. 
The last panel shows the galaxies' deviation $\Delta\sigma$ (see text for definition)
from the $M_{\rm BH}-\sigma_*$ relation, in dependence of Eddington ratio. An
apparent correlation does no longer exist once the blue outliers (marked
with extra open squares) are removed.  
}
\label{divcorr}
\end{figure*}

\begin{figure}
\plotone{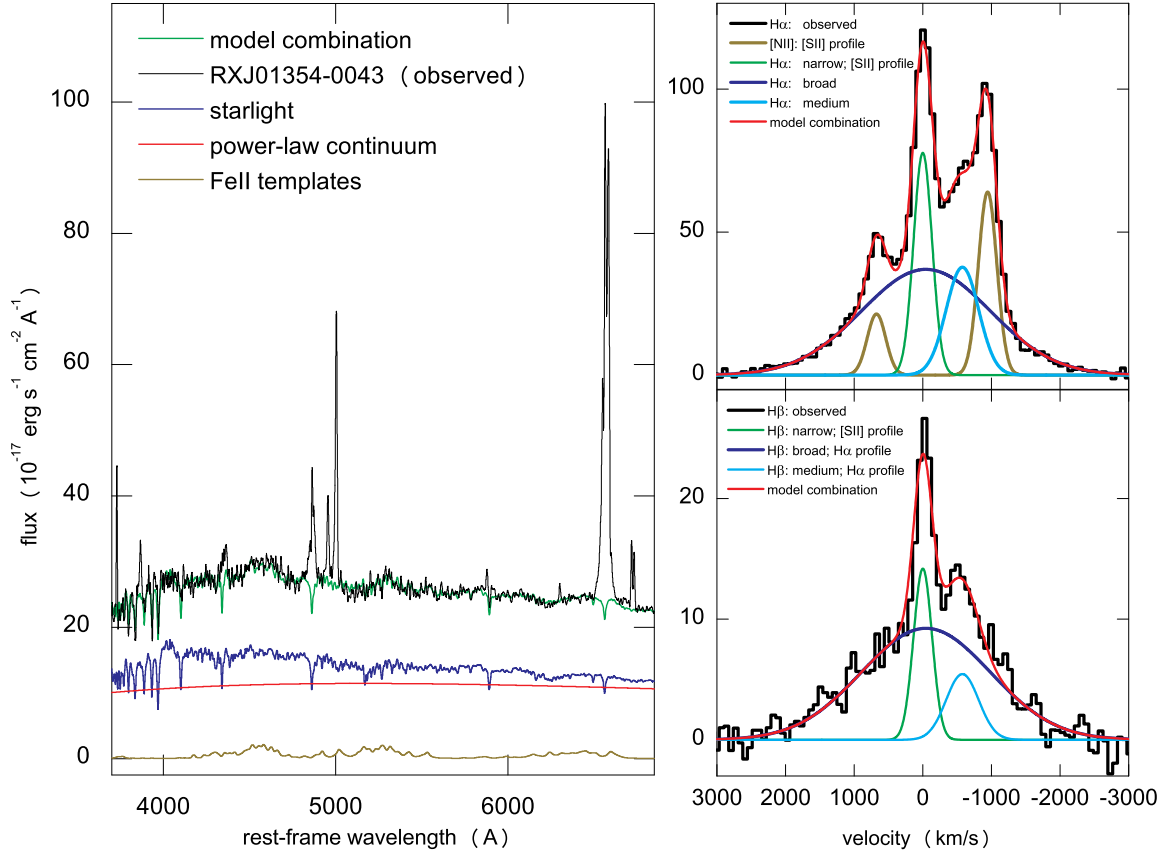} 
\caption{Left: SDSS spectrum of RXJ0135$-$0043, and component decomposition
into FeII complexes, a reddened AGN powerlaw continuum, and host galaxy
contribution as labeled in the figure. 
Strong absorption lines from the host galaxy are visible. 
Right: Zoom onto the H$\alpha$ and H$\beta$ lines, showing the presence of
two narrow cores of each Balmer line (green and light blue;
note that in this paper, we use positive velocities to indicated blueshifts,
negative values for redshifts), 
and a broad base (dark blue).
{\it{[See the electronic edition of the Journal for a colour version of this figure.]}}  
}
\label{decomp}
\end{figure}

\end{document}